\documentclass[conference]{IEEEtran}
\IEEEoverridecommandlockouts

\usepackage{cite}
\usepackage{amsmath,amssymb,amsfonts}
\usepackage{algorithmic}
\usepackage{algorithmic}
\usepackage{graphicx}
\usepackage{booktabs}
\usepackage{makecell}
\usepackage{textcomp}
\usepackage{xcolor}
\usepackage{marvosym}
\usepackage[caption=false,font=normalsize,labelfont=sf,textfont=sf]{subfig}
\usepackage{color}
\usepackage[colorlinks,citecolor=blue,urlcolor=blue,bookmarks=false,hypertexnames=true]{hyperref}

\def\BibTeX{{\rm B\kern-.05em{\sc i\kern-.025em b}\kern-.08em
    T\kern-.1667em\lower.7ex\hbox{E}\kern-.125emX}}

\begin{document}

\title{Robust Deep Joint Source-Channel Coding for Video Transmission over Multipath Fading Channel\\
\thanks{\textsuperscript{\Letter}Corresponding author: Yongsheng Liang (email: liangys@szu.edu.cn)}
}

\author{
\IEEEauthorblockN{Bohuai Xiao$^{1}$, Jian Zou$^{2,3}$, Fanyang Meng$^{4}$, Wei Liu$^{5}$, and Yongsheng Liang$^{1,3,}$\textsuperscript{\Letter}}

\IEEEauthorblockA{
$^{1}$College of Electronics and Information Engineering, Shenzhen University, Shenzhen, China\\
$^{2}$College of Applied Technology, Shenzhen University, Shenzhen, China\\ 
$^{3}$College of Big Data and Internet, Shenzhen Technology University, Shenzhen, China\\
$^{4}$Research Center of Networks and Communications, Pengcheng Laboratory, Shenzhen, China\\
$^{5}$Shenzhen Institute of Information Technology, Shenzhen, China}
Email: \{xiaobohuai2023, 2350414001\}@email.szu.edu.cn, mengfy@pcl.ac.cn, liuwei@sziit.edu.cn, liangys@szu.edu.cn
}

\maketitle

\begin{abstract}
To address the challenges of wireless video transmission over multipath fading channels, we propose a robust deep joint source-channel coding (DeepJSCC) framework by effectively exploiting temporal redundancy and incorporating robust innovations at the modulation, coding, and decoding stages. At the modulation stage, tailored orthogonal frequency division multiplexing (OFDM) for robust video transmission is employed, decomposing wideband signals into orthogonal frequency-flat sub-channels to effectively mitigate frequency-selective fading. At the coding stage, conditional contextual coding with multi-scale Gaussian warped features is introduced to efficiently model temporal redundancy, significantly improving reconstruction quality under strict bandwidth constraints. At the decoding stage, a lightweight denoising module is integrated to robustly simplify signal restoration and accelerate convergence, addressing the suboptimality and slow convergence typically associated with simultaneously performing channel estimation, equalization, and semantic reconstruction. Experimental results demonstrate that the proposed robust framework significantly outperforms state-of-the-art video DeepJSCC methods,
which achieves an average reconstruction quality gain of 5.13 dB under challenging multipath fading channel conditions\footnote{The code is available at https://github.com/MaximilienX/RVDJSCC}.
\end{abstract}


\section{Introduction}
Driven by the vision for sixth-generation (6G) wireless communication systems~\cite{b1}, 
emerging intelligent video-driven applications, 
such as virtual reality, telemedicine, and video conferencing,
are expected to impose unprecedented demands on network bandwidth, latency, and throughput. 
However, conventional communication systems based on separate source channel coding (SSCC) design~\cite{b2} 
have reached their respective performance limits in source compression and channel coding. 
Moreover, these systems suffer from the inherent cliff effect~\cite{b3}, 
leading to sharp performance degradation when channel conditions deteriorate below a certain threshold.
Consequently, traditional SSCC methods become inadequate to meet the future requirements of wireless video communications characterized by dynamic channel environments, bandwidth constraints, and limited power budgets.
Semantic communication, which aims to convey the meaning of information rather than raw bits, has long been envisioned as a promising paradigm to overcome the limitations of traditional communication systems. In recent years, its practical development has been significantly accelerated by advances in deep learning, enabling end-to-end semantic encoding and decoding that improve communication efficiency and robustness.

Among various semantic communication technologies, deep joint source channel coding (DeepJSCC) has attracted considerable research attention as a promising implementation of semantic communication~\cite{b4,b5}, offering a new avenue to overcome the performance bottlenecks of traditional SSCC approaches. 
Existing studies in DeepJSCC have tackled various core challenges inherent in its implementation, including variable bandwidth allocation strategies~\cite{b6} and optimal resource allocation techniques~\cite{b7}, significantly enhancing flexibility and performance. In addition, numerous DeepJSCC frameworks have been developed specifically for different types of information sources, such as images~\cite{b8,b9}, text~\cite{b7,b10}, and channel state information (CSI)~\cite{b11,b12}. 
Furthermore, recent research has integrated DeepJSCC with complementary communication techniques, including orthogonal frequency division multiplexing (OFDM)~\cite{b13}, non-orthogonal multiple access~\cite{b14}, and non-orthogonal superimposed pilot~\cite{b15}, to further enhance the robustness and efficiency of the system.

\begin{figure*}[!t]
\centering
\includegraphics[width=7.4in]{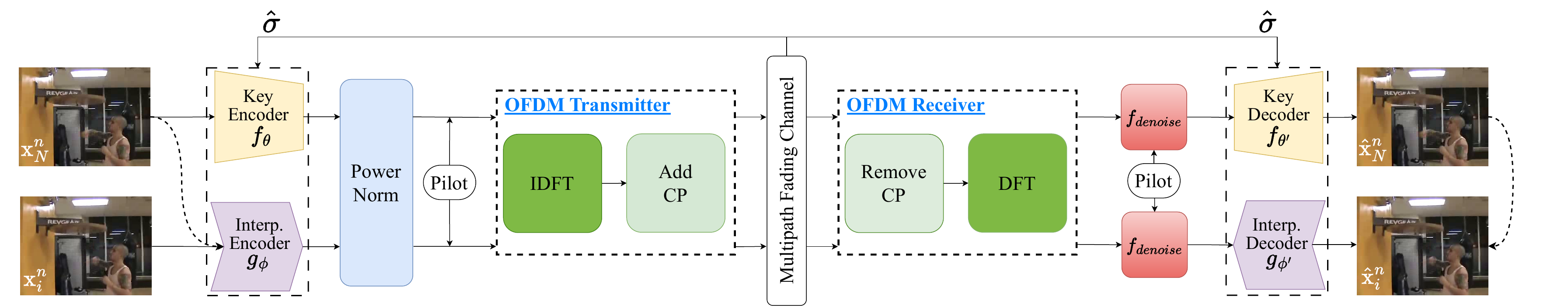}
\caption{Architecture of the proposed robust OFDM-based DeepJSCC framework for video transmission. It consists of two parallel coding paths: key-frame and interpolation-frame codecs. Dashed lines indicate reference frame inputs for interpolation coding and reconstruction.}
\label{fig:overview}
\end{figure*}

In particular, video data poses unique challenges for DeepJSCC due to its temporal continuity and substantial inter-frame redundancy.
DeepWiVe~\cite{b16} extends the DeepJSCC framework specifically for video by employing autoencoders to learn and compress inter-frame redundancy, and dynamically allocating bandwidth according to the importance of frames within a group of pictures (GOP). 
However, DeepWiVe commonly employs residual-based methods, which insufficiently capture semantic and temporal correlations, limiting compression efficiency. 
Another notable work, termed deep video semantic transmission (DVST)~\cite{b17}, formulates the DeepJSCC design as an optimization problem aimed at simultaneously enhancing perceived video quality and machine vision task performance. 
By emphasizing semantic fidelity rather than bit-level reconstruction, DVST reflects a core advantage of semantic communication. 
Nevertheless, these approaches remain limited to simple additive white Gaussian noise (AWGN) channels, neglecting the more complex multipath fading environments encountered in practical scenarios. 
Hence, there remains an urgent need for a more robust video DeepJSCC framework tailored specifically for multipath fading channels, to meet the stringent requirements of future intelligent video communications.

\begin{figure*}[!t]
\centering
\includegraphics[width=6.8in]{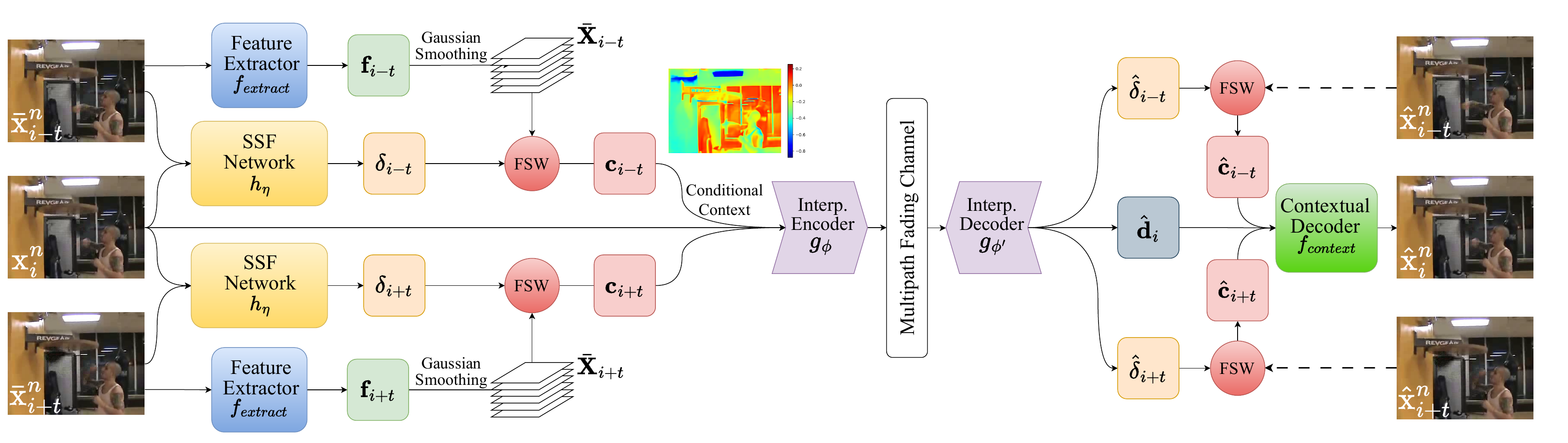}
\caption{Architecture of the interpolation network. Scale-space volume $\bar{\mathbf{X}}_{i\pm t}$ generated via Gaussian smoothing provides multi-scale features, warped by SSF $\boldsymbol{\delta}_{i\pm t}$ to obtain robust contextual conditions $\mathbf{c}_{i\pm t}$.
The dashed lines at the decoding end indicate the same generation process of $\bar{\mathbf{X}}_{i\pm t}$ as that at the encoding end.}
\label{fig:interp}
\end{figure*}

Motivated by the above challenges, we propose a robust DeepJSCC framework by effectively exploiting temporal redundancy and introducing dedicated innovations at the modulation, coding, and decoding stages. The main contributions are summarized as follows:
\begin{itemize}
\item 
We introduce OFDM at the modulation stage to combat frequency-selective fading caused by multipath propagation. By decomposing wideband signals into orthogonal sub-carriers that each experience flat fading, OFDM enhances signal robustness and supports implicit channel state learning through pilot embedding.
\item
We adopt a conditional context encoding mechanism at the coding stage to replace traditional residual-based methods. By leveraging multi-scale contextual features as conditions, this design captures richer temporal and semantic dependencies across frames, enabling more efficient compression of inter-frame redundancy under bandwidth and channel constraints.
\item 
We incorporate a lightweight denoising module at the decoding stage to reduce the complexity of long decoding chains that entangle channel equalization, estimation, and semantic reconstruction. By decoupling signal restoration from semantic decoding, this module streamlines the decoding process and improves convergence under noisy channel conditions.
\item 
Experimental results demonstrate that the introduction of OFDM modulation provides an average gain of \textbf{2.76} dB, 
the conditional contextual coding mechanism contributes an additional improvement of \textbf{1.8} dB, and the lightweight denoising module further enhances performance by \textbf{0.57} dB, cumulatively leading to an overall average reconstruction quality gain of \textbf{5.13} dB.
\end{itemize}

\section{System Model}
\label{sec:model}
We address the problem of wireless video transmission under stringent bandwidth constraints in multipath fading channels. 
As illustrated in Fig.~\ref{fig:overview},
the proposed system employs an OFDM-based DeepJSCC framework that directly maps video frames to complex-valued channel symbols. 
By integrating OFDM, the framework not only mitigates frequency-selective fading but also embeds pilot symbols for implicit channel estimation, thereby obviating the need for explicit CSI.

The multipath fading channel is parameterized by a channel transfer function:
\begin{equation}
\label{eq: 1}
\hat{\mathbf{y}}=\Upsilon(\mathbf{y})=h*\mathbf{y}+n,
\end{equation}
where $*$ denotes the convolution operation, 
$\mathbf{y}$ and $\hat{\mathbf{y}}$ represent channel input and output respectively,
$h\in\mathbb{C}^{L}$ represents the sample space channel impulse response, and $L$ is the number of multipaths.
$n\sim\mathcal{CN}(0,\sigma^2I)$ denotes the AWGN
with variance $\sigma^2$ and $I$ is the identity matrix.
Each path experiences independent Rayleigh fading satisfying $h_l\sim\mathcal{CN}(0,\sigma_l^2)$ for $l=0,1,\ldots,L-1$.
The variance $\sigma_l^2$ of each path follows the exponential decay, that is, $\sigma_l^2=\alpha_le^{-\frac{l}{\gamma}}$, where $\gamma$ represents the delay and $\alpha_l$ is a normalization coefficient. The sum of variances of all paths equals to 1, that is, $\sum_{l=0}^{L-1}\sigma_l^2=1$.
To simulate a dynamic fading environment, the channel gain $h$ is assumed to vary independently for each frame within a GoP.

Let $\mathbf{X}=\{\mathbf{X}^n\}_{n=1}^T$ be a video sequence composed of $T$ GoPs, where each GoP $\mathbf{X}^{n}=\{\mathbf{x}_{1}^{n},\ldots,\mathbf{x}_{N}^{n}\}$ contains $N$ frames. Each frame $\mathbf{x}_{i}^{n}\in\mathbb{R}^{H\times W\times3}$, $\forall i\in[1,N]$
is an 8-bit RGB image with three color channels. 
We design an encoding function
$E:\mathbb{R}^{H\times W\times3}\mapsto\mathbb{C}^{k}$
maps a video frame to the compressed representation
$\mathbf{z}=E(\mathbf{x})\in\mathbb{C}^{k}$, 
$k$ is the length of the compressed representation,
$\mathbf{z}$ is then reshaped into frequency domain symbols $\mathbf{Y}\in\mathbb{C}^{M\times N_s\times N_c}$, which are then fed into the OFDM transmitter. The pilot symbols $\mathbf{Y}_p\in\mathbb{C}^{M\times N_p\times N_c}$ are known to both the transmitter and receiver. $M$ is the number of OFDM packets, each packet contains $N_p$ pilot symbols and $N_s$ information symbols.
We apply inverse discrete Fourier transform (IDFT) to each OFDM symbol and append the cyclic-prefix (CP).
The OFDM transmit signal $\mathbf{y}\in\mathbb{C}^{M(N_s+N_p)(N_c+L_{cp})}$ propagates through the multipath channel defined by Eqn. \ref{eq: 1}.
Here, $N_c$ and $L_{cp}$ denote the number of subcarriers and the length of CP.
When the receiver obtains the channel output $\hat{\mathbf{y}}$, it removes the CP, and applies DFT to produce the frequency domain pilots $\hat{\mathbf{Y}}_p$ and data symbols $\hat{\mathbf{Y}}$:
\begin{equation}
\label{eq: 2}
\hat{\mathbf{Y}}_p[i,m]=H[m]\mathbf{Y}_p[i,m]+W[i,m],
\end{equation}
\begin{equation}
\label{eq: 3}
\hat{\mathbf{Y}}[j,m]=H[m]\mathbf{Y}[j,m]+V[j,m],
\end{equation}
where $H$ denotes the channel frequency response for the $m$-th subcarrier, and both $W$ and $V$ denote noise samples.
Then a decoding function is designed to reconstructs the original video frame as $\hat{\mathbf{x}}=D(\hat{\mathbf{Y}},\hat{\mathbf{Y}}_p,\mathbf{Y}_p)$ 
from the channel output given $\hat{\mathbf{Y}}$, $\hat{\mathbf{Y}}_p$
and $\mathbf{Y}_p$.

Considering practical power constraints,
the data symbols $\mathbf{Y}$ are subject to an average power constraint $P$:
\begin{equation}
\frac{1}{MN_s N_c}\mathbb{E}_\mathbf{Y}\left[||\mathbf{Y}||_2^2\right]\leq P,
\end{equation}
where the expectation is over the distribution of the encoder output.

We define the bandwidth constraint as
\begin{equation}
\rho=\frac{M(N_s+N_p)(N_c+L_{cp})}{3HW}.
\end{equation}

\section{Proposed Method}
In this section, we present the proposed robust DeepJSCC scheme for wireless video transmission. First, we describe the overall architecture in which key and interpolation frames are encoded via a contextual modeling approach to exploit temporal redundancy.
Then we introduce a dedicated denoising module that separates channel equalization and signal recovery from the decoding process, which effectively reduces the decoder’s complexity.

\subsection{Framework of Joint Source-Channel Video Coding}
The proposed DeepJSCC framework is based on~\cite{b16},
the video coding framework is divided into two parts: the key frame encoder/decoder
$(f_{\boldsymbol{\theta}},f_{\boldsymbol{\theta}^{\prime}})$, parameterized by 
$(\boldsymbol{\theta},\boldsymbol{\theta}^{\prime})$
and the interpolation frame encoder/decoder $(g_{\boldsymbol{\phi}},g_{\boldsymbol{\phi}^{\prime}})$, parameterized by
$(\boldsymbol{\phi},\boldsymbol{\phi}^{\prime})$.
For the $n$th GoP, $\mathbf{X}^n=\{\mathbf{x}_1^n,\ldots,\mathbf{x}_N^n\}$. The last frame $\mathbf{x}_N^n$ is called the key frame and is processed and transmitted by key frame encoder.

The remaining frames $\mathbf{x}_i^n,i=1,\ldots,N-1$ are encoded by interpolation encoder with respect to two reference frames
$(\bar{\mathbf{x}}_{i-t}^n,\bar{\mathbf{x}}_{i+t}^n)$
that are $t$ frames away from the current one.
We define $\bar{\mathbf{x}}_0^n=\bar{\mathbf{x}}_N^{n-1}$, and assume that GoPs are processed sequentially with prior decoded frames available as references.
As shown in Fig.~\ref{fig:overview},
both the key and interpolation encoders integrate an Attention Feature (AF) module~\cite{b7}
that dynamically reweights feature channels based on instantaneous SNR, which allows a single model to adapt across varying channel conditions. This cross-SNR adaptability streamlines training and improves efficiency by dynamically allocating representational capacity according to channel quality.

The key frame encoder 
$f_{\boldsymbol{\theta}}:\mathbb{R}^{H\times W\times3}\mapsto\mathbb{C}^{k}$ maps the key frame $\mathbf{x}_N$ 
to $\mathbf{z}_N$:
\begin{equation}
\mathbf{z}_N=f_{\theta}(\mathbf{x}_N,\hat{\sigma}^2),
\end{equation}
where $\hat{\sigma}^2$ is the estimated channel noise power.
The compressed representation $z_N$ is then reshaped into OFDM packets and, after pilot insertion, IDFT, and CP addition, to obtain the transmit signal $\mathbf{y}_N\in\mathbb{C}^{M(N_s+N_p)(N_c+L_{cp})}$.
The signal passes through the multipath fading channel and is then processed by the OFDM receiver as shown in Eqns.~\ref{eq: 2} and~\ref{eq: 3}.
The key frame decoder $f_{\boldsymbol{\theta}^{\prime}}:\mathbb{C}^{M(N_s+N_p+1)N_c}\mapsto\mathbb{R}^{H\times W\times3}$ reconstructs the key frame $\hat{\mathrm{x}}$ using the OFDM receiver output $(\hat{\mathbf{Y}},\hat{\mathbf{Y}}_p)_N$ and the original pilot symbols $\mathbf{Y}_p$:
\begin{equation}
\label{eq: 7}
\hat{\mathrm{x}}_N=f_{\boldsymbol{\theta'}}
((\hat{\mathbf{Y}},\hat{\mathbf{Y}}_p)_N,\mathbf{Y}_p,\hat{\sigma}^2).
\end{equation}

To further reduce inter-frame redundancy and improve video coding efficiency, we introduce our contextual feature coding mechanism for interpolation frames, which leverages feature-domain context in place of traditional pixel-domain residual coding to more effectively capture temporal correlations in the next subsection.

\subsection{Conditional Context Coding Mechanism}
Inspired by the works on deep contextual video compression~\cite{b18,b19,b20}, we adopt the idea of conditional context coding to enhance the compression of inter-frame redundancy. Specifically, we move beyond traditional pixel-domain residual coding, which uses a simple subtraction-based operation to model temporal dependencies, and instead leverage feature-domain context to model these dependencies in a more expressive way. By using contextual features from reference frames, we aim to achieve a more efficient compression of interpolation frames.
The architecture of the interpolation frame encoding network is shown as Fig.~\ref{fig:interp}.

We use the same network architecture 
$h_{\boldsymbol{\eta}}:\mathbb{R}^{H\times W\times6}\mapsto\mathbb{R}^{H\times W\times3}$ as in~\cite{b16}, to estimate the scaled space flow (SSF)~\cite{b21}
$(\boldsymbol{\delta}_{i-t},\boldsymbol{\delta}_{i+t})$ , which models the multi-scale motion information between the reference frames $(\mathbf{\bar{x}}_{i-t},\mathbf{\bar{x}}_{i+t})$ and the interpolation frame $\mathbf{x}_i$ to be encoded:
\begin{equation}
\boldsymbol{\delta}_{i\pm t}=
h_{\boldsymbol{\eta}}(\mathbf{x}_i,\mathbf{\bar{x}}_{i\pm t}).
\end{equation}

To capture richer contextual information, particularly the dependencies between pixels at different positions across frames, which traditional residual encoding methods are unable to model effectively, we map a reference frame $\bar{\mathbf{x}}_{i-t}$  to a high-dimensional feature $\mathbf{f}_{i-t}$ through a feature extractor $f_{extract}:\mathbb{R}^{H\times W\times3}\mapsto\mathbb{R}^{H\times W\times C_{f}}$.
Then we generate a fixed-resolution scale-space volume $\bar{\mathbf{X}}_{i-t}\in\mathbb{R}^{H\times W\times(V+1)}$ from the reference frame’s high-dimensional feature $\mathbf{f}_{i-t}$,
in order to further apply feature-space warping (FSW) in feature domain:
\begin{equation}
\label{eq: 9}
\begin{split}
\bar{\mathbf{X}}_{i-t} = & \, [\mathbf{f}_{i-t}, \mathbf{f}_{i-t}*G(\sigma_0), \cdots, \\
                          & \, \mathbf{f}_{i-t}*G(2^{V-1}\sigma_0) ],
\end{split}
\end{equation}
where $\mathbf{f}_{i-t}*G(\sigma_0)$ denotes the convolution of $\mathbf{f}_{i-t}$ and a Gaussian smoothing kernel with scale $\sigma_0$, $V$ is the number of levels in the scale-space volume.

The SSF $\boldsymbol{\delta}_{i-t}$ is applied to the scale-space volume $\bar{\mathbf{X}}_{i-t}$ to perform an FSW operation,
obtaining the feature domain context:
\begin{equation}
\label{eq: 10}
\mathbf{c}_{i-t}=\mathrm{FSW}(\bar{\mathbf{X}}_{i-t},\boldsymbol{\delta}_{i-t}),
\end{equation}
and the other reference frame $\bar{\mathbf{x}}_{i+t}$ follows the same process to obtain the context $\mathbf{c}_{i+t}$.

The contexts $(\mathbf{c}_{i-t},\mathbf{c}_{i+t})$ generated from the reference frames are used as conditions, the interpolation encoder $g_{\boldsymbol{\phi}}:\mathbb{R}^{H\times W\times(2C_f+3)}\mapsto\mathbb{C}^k$ maps the interpolation frame combined with the condition to the compressed representation, which is given by
\begin{equation}
\mathbf{z}_i=g_{\boldsymbol{\phi}}(\mathbf{x}_i,\mathbf{c}_{i-t},\mathbf{c}_{i+t},\hat{\sigma}^2),
\end{equation}
after being modulated through the same OFDM process as the key frame and transmitted through the multipath channel, given the receiver output $(\hat{\mathbf{Y}},\hat{\mathbf{Y}}_p)_i$, the interpolation decoder estimates the SSF 
$(\hat{\boldsymbol{\delta}}_{i-t},\hat{\boldsymbol{\delta}}_{i+t})$ 
and a preliminary decoded representation
$\hat{\mathbf{d}}_i$:
\begin{equation}
\label{eq: 12}
(\hat{\mathbf{d}}_i,
\hat{\boldsymbol{\delta}}_{i-t},
\hat{\boldsymbol{\delta}}_{i+t})
=g_{\boldsymbol{\phi}^{\prime}}
((\hat{\mathbf{Y}},\hat{\mathbf{Y}}_p)_i,
\mathbf{Y}_p,
\hat{\sigma}^2).
\end{equation}

Finally, we construct the interpolation frame utilizing the structure designed symmetrically like the encoding end, after obtaining conditional contexts $(\hat{\mathbf{c}}_{i-t},\hat{\mathbf{c}}_{i+t})$ as Eqns. \ref{eq: 9} and \ref{eq: 10}:
\begin{equation}
\label{eq: 13}
\hat{\mathbf{x}}_i=f_{context}(\hat{\mathbf{d}}_i,\hat{\mathbf{c}}_{i-t},\hat{\mathbf{c}}_{i+t}),
\end{equation}
where $f_{context}$ is a contextual decoder proposed in~\cite{b18}. 

\subsection{Denoising-Aided Joint Decoding}

\begin{figure}[!t]
\centerline{\includegraphics[width=0.38\textwidth]{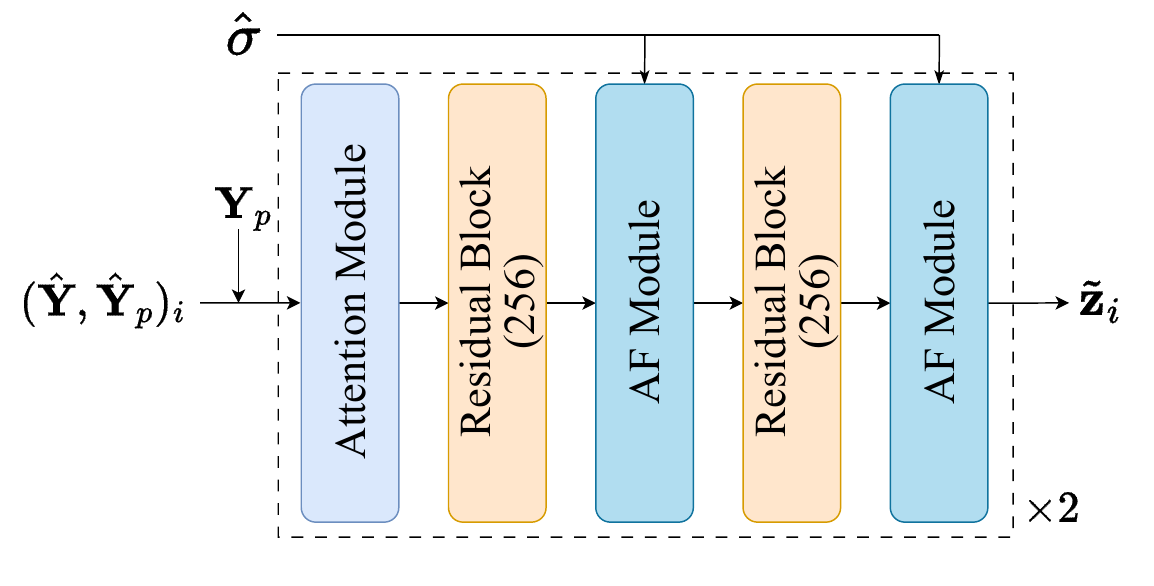}}
\caption{Architecture of the denoising module $f_{denoise}$.}
\label{fig:denoise}
\end{figure}


In conventional OFDM-based JSCC decoders, the entire received signal, comprising both pilot and information components, is processed by a single deep neural network as described in Eqns. \ref{eq: 7}, \ref{eq: 12}, and \ref{eq: 13}. This unified processing chain excessively burdens the decoder, requiring it to simultaneously handle tasks of channel estimation, equalization, denoising, and semantic reconstruction. Consequently, this prolonged sequence of tasks results in slower convergence and suboptimal reconstruction quality.

To mitigate these challenges, we propose to decouple the denoising step from the decoding process into two distinct stages. Initially, a lightweight denoising module, illustrated in Fig. \ref{fig:denoise}, leverages the pilot-derived channel information to preprocess and refine the raw composite signal.
Subsequently, this denoised signal is input into a dedicated deep decoder that is exclusively tasked with reconstructing the original video frame.
This modular design effectively shortens the processing chain, significantly reduces the decoder's learning complexity, accelerates model convergence, and ultimately enhances the quality of the reconstructed video frames.

Specifically, the denoising network processes the OFDM receiver output $(\hat{\mathbf{Y}},\hat{\mathbf{Y}}_p)_i$
to produce a clean compressed representation $\tilde{\mathbf{z}}_i$:
\begin{equation}
\tilde{\mathbf{z}}_i=f_{denoise}
((\hat{\mathbf{Y}},\hat{\mathbf{Y}}_p)_i,\mathbf{Y}_p,\hat{\sigma}^2),
\end{equation}
after denoising, the key and interpolation decoders map the clean latent vector to reconstructed frame respectively.

For the key decoder, it is simplified as:
\begin{equation}
\hat{\mathbf{x}}_N=f_{\boldsymbol{\theta}^{\prime}}(\tilde{\mathbf{z}}_N,\hat{\sigma}^2).
\end{equation}

Similarly, the interpolation decoder is given by
\begin{equation}
(\hat{\mathbf{d}}_i,
\hat{\boldsymbol{\delta}}_{i-t},
\hat{\boldsymbol{\delta}}_{i+t})=
g_{\boldsymbol{\phi}^{\prime}}(\tilde{\mathbf{z}}_i,\hat{\sigma}^2),
\end{equation}
and the subsequent process remains unchanged as shown in Eqns. \ref{eq: 9}, \ref{eq: 10}, and \ref{eq: 13}.

\subsection{Optimization Goal}
The training loss in our proposed method consists of two main components. First, the mean-squared-error (MSE) loss between the latent variable before the channel and its denoised counterpart, which ensures effective denoising. Second, the MSE loss between the original and the reconstructed video frames, ensuring high-quality video reconstruction. These two MSE losses are weighted and combined to form the total training objective, for the training video sequence $\mathbf{X}$, the loss function is as follows
\begin{equation}
L(\mathbf{X},\hat{\mathbf{X}})=\frac{1}{TN}\sum_{n=1}^T\sum_{i=1}^N\mathrm{MSE}(\mathbf{x}_i^n,\hat{\mathbf{x}}_i^n)\ + \lambda\cdot\mathrm{MSE}(\mathbf{z}_i^n,\tilde{\mathbf{z}}_i^n),
\end{equation}
the weight parameter $\lambda$ controls the relative importance of the reconstruction and the denoising loss. In this paper, $\lambda=0.7$.

\section{Numerical Results}

\subsection{Training Details}
We utilize the UCF101 dataset~\cite{b22}, a widely-used benchmark containing over 13,000 videos across 101 human action categories, to train the proposed DeepJSCC framework for video transmission. For the OFDM and multipath fading channel, we set the parameters as follows: $N_p=2,\ N_s=12,\ N_c=256,\ L_{cp}=16,\ L=8,\ \gamma=4$, and $P=1$. Additionally, we follow the GoP-based interpolation structure described in~\cite{b15}, with a GoP size $N=4$, where $t=2$ for $i=2$ and $t=1$ for $i=1,3$. To maintain consistent compressed representation length $k$ per GoP, we set $M=6$ for key frames and $M=2$ for interpolation frames. Due to hardware memory constraints (NVIDIA Tesla V100 GPU with 16GB memory), the training batch size is limited to 1.

We employ two metrics to evaluate the average quality of reconstructed video frames: peak signal-to-noise ratio (PSNR) and multi-scale structural similarity index (MS-SSIM), defined as follows: \begin{equation} \mathrm{PSNR}(\mathbf{X},\hat{\mathbf{X}})=\frac{1}{TN}\sum_{n=1}^{T}\sum_{i=1}^{N}10\log_{10}\left(\frac{255^2}{l_{\mathrm{PSNR}}(\mathbf{x}_i^n,\hat{\mathbf{x}}_i^n)}\right), \end{equation}
\begin{equation} \mathrm{MS\text{-}SSIM}(\mathbf{X},\hat{\mathbf{X}})=\frac{1}{TN}\sum_{n=1}^{T}\sum_{i=1}^{N}\left[1-l_{\mathrm{MS\text{-}SSIM}}(\mathbf{x}{i}^{n},\hat{\mathbf{x}}{i}^{n})\right]. \end{equation}

Moreover, we adopt an early stopping strategy to mitigate overfitting and a learning-rate scheduler to ensure stable convergence. Specifically, the initial learning rate is set to $1\times10^{-4}$ and dynamically adjusted by a factor of 0.8 if the validation error does not improve for 4 consecutive epochs. Training is terminated if validation error stagnates for 8 epochs. The learning rate at epoch $n$ is thus defined as:
\begin{equation}
lr_{n} = 
\begin{cases}
0.8 \times lr_{n-1}, & \text{if } \text{bad epochs} \ge 4,\\
lr_{n-1}, & \text{otherwise}.
\end{cases}
\label{eq:lr_schedule}
\end{equation}

\subsection{Simulation Results}

To validate the effectiveness of the proposed framework under multipath fading channel conditions, we conduct evaluations focusing on (1) performance analysis and (2) complexity analysis.

\begin{figure*}[!t]
	\centering
	\begin{minipage}{0.329\linewidth}
		\centering
		\includegraphics[width=1.0\linewidth]{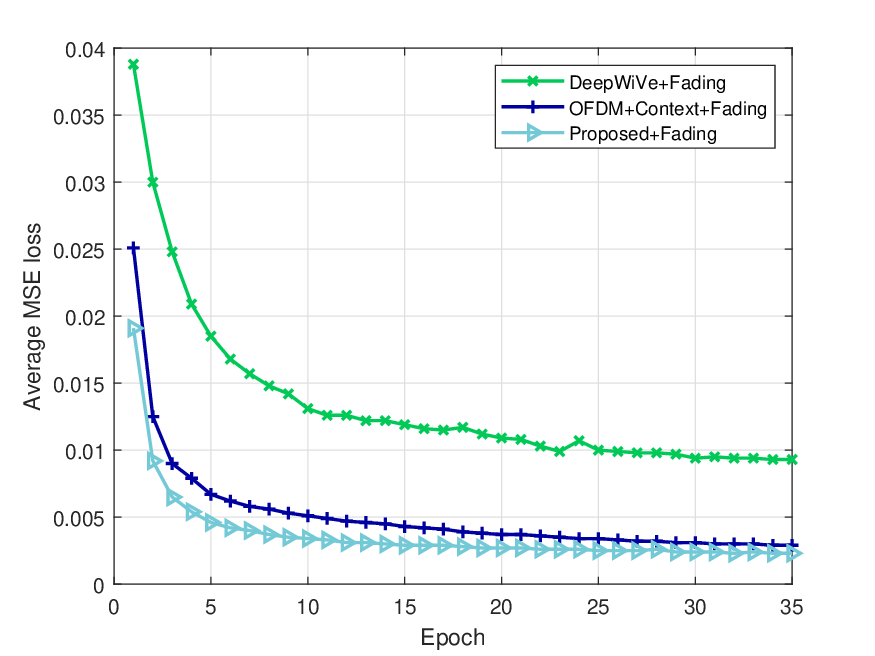}
		\caption{MSE loss curves, validating that the denoising module accelerates convergence.}
		\label{fig:exp2}
	\end{minipage}
	\begin{minipage}{0.329\linewidth}
		\centering
		\includegraphics[width=1.0\linewidth]{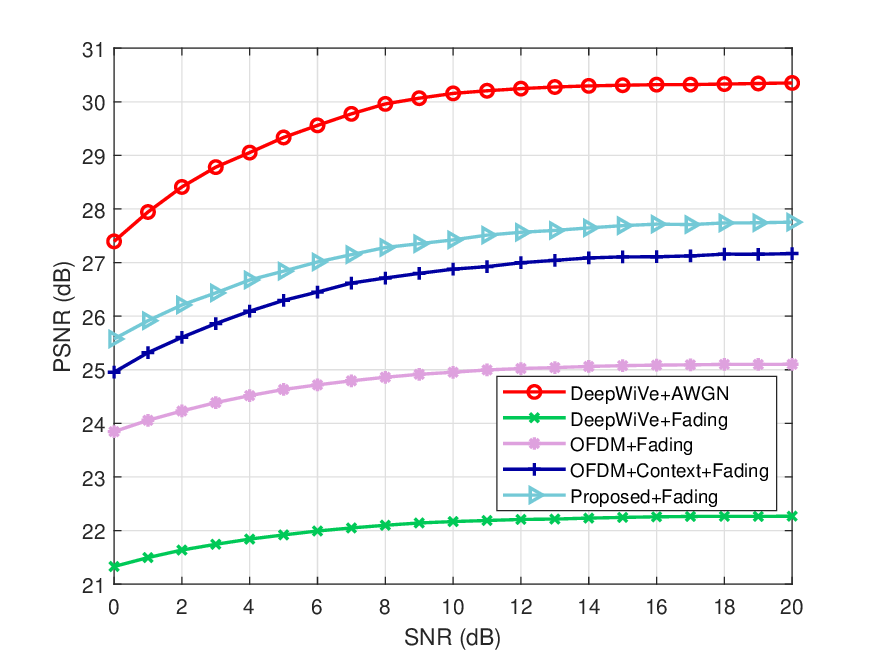}
		\caption{PSNR comparison of the proposed framework to the baseline and its ablated variants.}
		\label{fig:exp3}
	\end{minipage}
	\begin{minipage}{0.329\linewidth}
		\centering
		\includegraphics[width=1.0\linewidth]{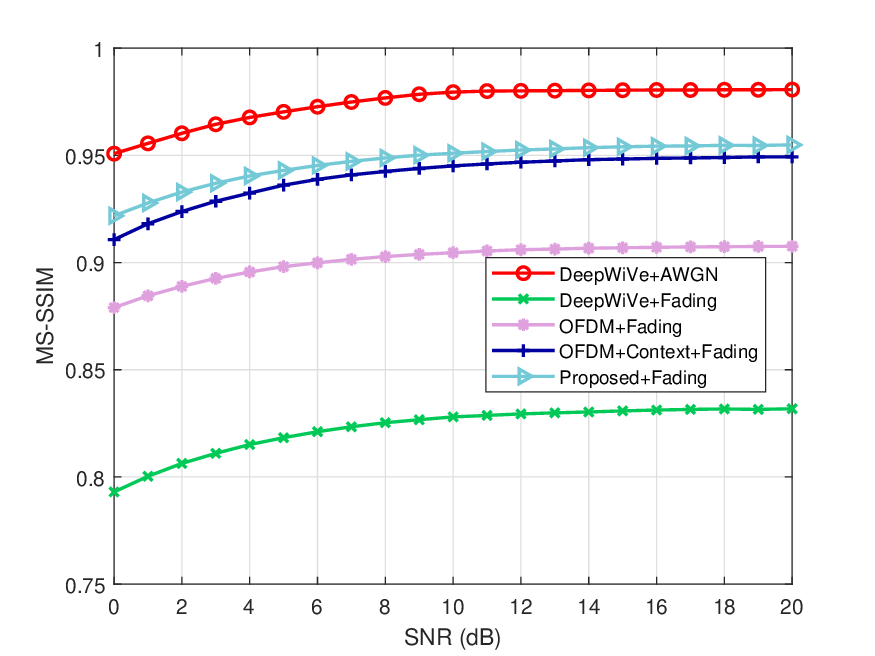}
		\caption{MS-SSIM comparison of the proposed framework to the baseline and its ablated variants.}
		\label{fig:exp4}
	\end{minipage}
\end{figure*}

\emph{1) Performance Analysis:}
Fig.~\ref{fig:exp2} illustrates the MSE loss curves of three configurations: baseline ``DeepWiVe+Fading'', ablation variant OFDM+Context+Fading''(without the denoising module), and the proposed framework ``Proposed+Fading''. The proposed framework exhibits the fastest convergence due to the integrated denoising module, simplifying the decoding task by decoupling channel estimation, equalization, and semantic reconstruction.

In Fig.~\ref{fig:exp3}, we present the PSNR performance of different
model configurations across varying SNR conditions.
``DeepWiVe+AWGN'' achieves an average PSNR of 29.68 dB, representing an ideal upper bound without fading. 
However, under multipath fading, the ``DeepWiVe+Fading'' experiences substantial degradation, reducing its PSNR to 22.03 dB.

An ablation variant ``OFDM+Fading''(without the conditional contextual coding mechanism and denoising module), 
which introduces OFDM technology at the modulation stage,
significantly improves the PSNR to 24.79 dB, clearly demonstrating its robustness in combating multipath fading by decomposing wideband signals into orthogonal frequency-flat sub-channels and implicitly estimating channel states via embedded pilots.
Another ablation variant ``OFDM+Context+Fading''(without the denoising module),
which integrates the conditional contextual coding mechanism at the encoding stage,
further enhances PSNR performance to 26.59 dB.
Notably, the benefit of conditional contextual coding mechanism becomes more prominent at higher SNR regimes. 
This is because the mechanism extracts finer-grained inter-frame redundancy features,
preserving richer semantic content at the same compression rate.
Hence, under better channel conditions, such semantic details
significantly enhance reconstruction quality.

Finally, the complete proposed framework achieves the highest average PSNR of 27.16 dB. These results validate that each module provides complementary contributions, collectively robustifying video reconstruction quality in realistic multipath fading environments.Consistent observations can also be made from Fig.~\ref{fig:exp4}, where the MS-SSIM metric aligns with the PSNR trends observed in the ablation studies, further confirming the robustness and enhanced perceptual quality achieved by the proposed DeepJSCC framework.

\begin{table}[!t]
\begin{center}
\tabcolsep=0.12cm
\renewcommand\arraystretch{1.5}
\caption{Computational Complexity and Delay Analysis.}
\label{Table I}
\begin{tabular}{ c  c  c}
\hline
\hline
Methods & Params ($ \times {10^7}$) & FLOPs ($ \times {10^{11}}$)\\
\hline
Proposed & 7.375 & 5.054\\
OFDM+Context & 6.773 & 5.037\\
OFDM & 6.713 & 3.222\\
DeepWiVe & 6.713 & 3.222\\
\hline
\hline 
\end{tabular}
\end{center}
\end{table}

\emph{2) Complexity Analysis:}
As shown in Table~\ref{Table I},
the proposed robust DeepJSCC video transmission framework has a size of 73.8 MB and requires 505.4 G FLOPs of computation. Due to the incorporation of OFDM, which introduces pilot symbols and CP redundancy without additional network parameters, the channel occupancy experiences a slight increase of 24\% compared to the original architecture~\cite{b16}, even though the compression rate remains unchanged. 
Notably,
the baseline model originally includes network designs specifically intended for bandwidth allocation.
Thus, to maintain a fair comparison, we remove these redundant parameters from the baseline model.

\section{Conclusion}
A robust DeepJSCC framework is proposed for wireless video transmission over multipath fading channels. The method effectively addresses frequency-selective fading without explicit channel estimation by integrating pilot-based OFDM.
In addition, a conditional contextual coding mechanism with multi-scale Gaussian warped features better exploits inter-frame redundancy than conventional residual-based techniques. Lastly, a lightweight denoising module at the decoder simplifies the overall task and accelerates convergence. Experimental results confirm that this approach surpasses state-of-the-art video DeepJSCC frameworks under multipath fading conditions. Future work will explore advanced entropy models for fine-grained, data-driven bandwidth allocation based on inter-frame information, further enhancing system robustness and efficiency.


\section*{Acknowledgment}

This research was supported by the National Natural Science Foundation of China (Grant No. 62031013), the Guangdong Province Key Construction Discipline Scientific Research Capacity Improvement Project (Grant No. 2022ZDJS117), and the Shenzhen Technology Research Key Project (Grant No. JSGG20211029095545002).

\end{document}